\documentclass[a4paper]{article}
\usepackage{epsfig,amsmath,amsfonts,amssymb,setspace,multirow,textcomp}
\usepackage[T1]{fontenc}
\usepackage[sort&compress,numbers]{natbib}

\makeatletter
\renewcommand{\@oddhead}{\textit{Advances in Astronomy and Space Physics} \hfil}
\renewcommand{\@evenfoot}{\hfil \thepage \hfil}
\renewcommand{\@oddfoot}{\hfil \thepage \hfil}
\makeatother

\renewenvironment{thebibliography}[1]{\begin{oldthebibliography}{#1}\setlength{\parskip}{0ex}\setlength{\itemsep}{0ex}}{\end{oldthebibliography}}

\usepackage{xspace}
\usepackage{color}

\newcommand{\xmm}{\textit{XMM-Newton}\xspace}
\newcommand{\chan}{\textit{Chandra}\xspace}
\newcommand{\suza}{\textit{Suzaku}\xspace}
\newcommand{\dm}{{\textsc{dm}}} 
\renewcommand{\S}{\mathcal{S}} 
\newcommand{\numsm}{$\nu$MSM} 

\usepackage{tabularx}

\usepackage{booktabs} 
\newcommand {\otoprule }{\midrule [\heavyrulewidth]} 
\usepackage{amsmath,amsfonts,amssymb}

\begin{document}
\fontsize{11}{11}\selectfont 
\title{Observation of the new emission line at $\sim$3.5 keV in X-ray spectra \\ 
of galaxies and galaxy clusters}
\author{\textsl{D.\,A.~Iakubovskyi$^{1,2}$}}
\date{\vspace*{-6ex}}
\maketitle
\begin{center} {\small $^{1}$Discovery Center, Niels Bohr Institute, Blegdamsvej 17, Copenhagen, Denmark\\
$^{2}$Bogolyubov Institute of Theoretical Physics, Metrologichna Str. 14-b, 03680, Kyiv, Ukraine\\
{\tt iakubovskyi@nbi.ku.dk}}
\end{center}

\begin{abstract}
The detection of an unidentified emission line in X-ray spectra of cosmic objects 
would be a `smoking gun' signature for particle physics beyond the Standard Model. 
More than a decade of its extensive searches results in several narrow faint emission lines
reported at 3.5, 8.7, 9.4 and 10.1~keV. The most promising
of them is the emission line at $\sim$3.5~keV reported in spectra of 
several nearby galaxies and galaxy clusters. Here I summarize its up-to-date status, 
overview its possible interpretations, including an intriguing connection with radiatively decaying 
dark matter, and outline future directions for its studies.\\[1ex]
{\bf Key words:} X-rays: general, dark matter, line: identification.
\end{abstract}

\twocolumn[]

\section*{\sc introduction}\label{sec:previous-searches}

\indent \indent The origin of \emph{dark matter} -- the major (yet of unknown origin) 
gravitating substance in the Universe \cite{Zwicky:33,Sarazin:86,Evrard:95,Einasto:99,Bergstrom:00,
Corbelli:03,Refregier:03,Dekel:05,Massey:07,Gilmore:07a,Noordermeer:07,Fu:07,Coccato:08,
Einasto:09,Rozo:09,Reid:09,Chemin:09,Corbelli:09,Massey:10,Tinker:11,Roos:12,Frenk:12,Hinshaw:12,
Walker:13book,Planck2015-XIII} -- still has to be revealed. If dark matter is made of elementary particles,
the corresponding particle should be massive (to form over-densities in process of gravitational collapse), 
long-lived (to be stable for at least the age of the 
Universe) and neutral with respect to strong and electromagnetic interactions 
(to be sufficiently `dark'). The only known massive, long-lived and neutral particles are the usual (left-handed) 
neutrinos, but they are too light to form small dark matter haloes \cite{Tremaine:79,White:83}.
As a result, the hypothesis of dark matter particle implies an extension of the Standard Model of particle 
physics. 
Dozens of the Standard Model extensions proposed so far to contain a valid dark matter particle candidate.
However, as Fig.~1 of \cite{Gardner:13} demonstrates, 
the masses of dark matter particle candidates and their interaction strengths 
with Standard Model particles cover a huge region of parameter space. This results in a large variety of
observational methods developed to search for dark matter particles.

The specific example considered in this review is \emph{radiatively decaying dark matter}. 
If a dark matter particle interacts with electrically charged particles, it may
\footnote{The widely-known 
examples where this is \emph{not} the case are dark matter particles are the \emph{lightest} particles 
holding a new quantum number 
\emph{conserved} by the Standard Model interactions, such as R-parity for super-symmetric models, Kaluza-Klein 
number for extra dimensions, etc.. In this case, dark matter decays are strictly forbidden 
by the special structure of the theory, 
and the main astrophysical effect for dark matter particles is their \emph{annihilation}
with their antiparticles.}
possess a radiative 
decay channel. If a non-relativistc dark matter particle decays to a photon and another particle, 
slight ($v/c \lesssim 5\times 10^{-3}$) Doppler broadening due to non-zero 
velocities of dark matter particles in halos would cause a narrow \emph{dark matter decay line}.
Such a decay line possesses several specific features allowing to robustly
distinct it from emission lines of astrophysical origin (see e.g. \cite{Dere:97,Smith:01}) 
or from instrumental line-like features:
\begin{itemize}
 \item its position in energy is solely determined by the mass of dark matter particle 
 and the redshift of dark matter halo 
 (i.e. if one neglects the mass of the other decay product, the line position is 
 $\frac{m_\dm c^2}{2(1+z)}$), having 
 different scaling with halo redshift $z$ compared with instrumental line-like features; 
 \item its intensity is proportional to \emph{dark matter column density}
 $\S_\dm=\int \rho_\dm d\ell$; 
 due to different 3D distributions of dark and visible matter, comparison
 of the new line intensity within the given object and among different objects would allow 
 to choose between its decaying dark matter and astrophysical origins; 
 \item it is broadened with the characteristic velocity of dark matter
 different from that of visible matter.
\end{itemize}  
The above-mentioned characteristics allow to \emph{directly detect the radiatively decaying dark matter
relying on astrophysical measurements}. This motivates the extensive search for new lines in X-ray spectra 
of cosmic objects proposed about 15 years ago \cite{Dolgov:00,Abazajian:01a,Abazajian:01b}, 
see Table~\ref{tab:bounds-summary}.
An example is the analysis of the line candidate at $\sim$2.5~keV initially 
reported by \cite{Loewenstein:09} in X-ray spectrum of the Willman~1 dwarf spheroidal at 2.5$\sigma$ 
level. Further non-observation of this line candidate in central part 
and outskirts of Andromeda galaxy, Fornax and Sculptor dwarf spheroidal galaxies \cite{Boyarsky:10a}  
excludes the decaying dark matter origin of the $\sim$2.5~keV signal at high significance level 
(above 14$\sigma$). This result is further strengthened
by the authors of \cite{Mirabal:10a} who reanalyzed the same observations of Willman~1 as \cite{Loewenstein:09}
(and did not find the $\sim$2.5~keV line)
and the authors of \cite{Mirabal:10b} who analysed another dwarf spheroidal, Segue~1. Finally, the authors 
of \cite{Loewenstein:12} ruled
out the dark matter origin of the $\sim$2.5~keV feature by looking at Willman~1 with better statistics. 
The probable origin of the $\sim$2.5~keV line, according to \cite{Boyarsky:10a}, 
is purely instrumental, being the result of under-modelling of the time-variable soft proton background 
(see e.g. \cite{Kuntz:08}) in some observations combined with an apparent dip 
at $\sim$2.5~keV in the effective area of existing X-ray instruments.

\begin{table*}
\begin{tabularx}{\textwidth}{llXc}
\toprule
Ref. & Object & Instrument & Cleaned exposure, ks  \\
\otoprule
\cite{Boyarsky:05} & Diffuse X-ray background & HEAO-1, \xmm/EPIC & 224, 1450  \\
\cite{Boyarsky:06b} & Coma, Virgo & \xmm/EPIC & 20, 40  \\
\cite{Boyarsky:06c} & Large Magellanic Cloud & \xmm/EPIC & 20  \\
\cite{Riemer-Sorensen:06a} & Milky Way & \chan/ACIS-S3 & Not specified  \\
\cite{Watson:06} & M31 (central $5'$) & \xmm/EPIC & 35  \\
\cite{Riemer-Sorensen:06b} & Abell~520 & \chan/ACIS-S3 & 67  \\
\cite{Boyarsky:06d} & Milky Way, Ursa Minor & \xmm/EPIC & 547, 7  \\
\cite{Abazajian:06b} & Milky Way & \chan/ACIS & 1500  \\
\cite{Boyarsky:06e} & 1E~0657-56 (``Bullet cluster'') & \chan/ACIS-I & 450  \\
\cite{Boyarsky:06f} & Milky Way & X-ray micro-calorimeter & 0.1  \\
\cite{Yuksel:07} & Milky Way & INTEGRAL/SPI & 5500  \\
\cite{Boyarsky:07a} & M31 (central $5-13'$) & \xmm/EPIC & 130  \\
\cite{Boyarsky:07b} & Milky Way & INTEGRAL/SPI & 12200 \\
\cite{Loewenstein:08} & Ursa Minor & \suza/XIS & 70  \\
\cite{Riemer-Sorensen:09} & Draco & \chan/ACIS-S & 32  \\
\cite{Loewenstein:09} & Willman~1 & \chan/ACIS-I & 100  \\
\cite{Boyarsky:10a} & M31, Fornax, Sculptor & \xmm/EPIC , \chan/ACIS & 400, 50, 162   \\
\cite{Mirabal:10a} & Willman~1 & \chan/ACIS-I & 100\\
\cite{Mirabal:10b} & Segue~1 & Swift/XRT & 5  \\
\cite{Borriello:11} & M33 & \xmm/EPIC & 20-30  \\
\cite{Watson:11} & M31 ($12-28'$ off-centre)& \chan/ACIS-I & 53 \\
\cite{Loewenstein:12} & Willman~1 & \xmm/EPIC & 60 \\
\cite{Kusenko:12} & Ursa Minor, Draco & \suza/XIS & 200, 200 \\
\cite{Iakubovskyi:13} & Stacked galaxies & \xmm/EPIC & 8500 \\
\cite{Horiuchi:13} & M31 & \chan/ACIS-I & 404 \\
\cite{Malyshev:14} & Stacked dSphs & \xmm/EPIC & 410 \\
\cite{Anderson:14} & Stacked galaxies & \xmm/EPIC, \chan/ACIS-I & 14600, 15000 \\
\cite{Tamura:14} & Perseus & \suza/XIS & 520\\
\cite{Horiuchi:15,Ng:15} & Milky Way & Fermi/GBM & 4600 \\
\cite{Sekiya:15} & Milky Way & \suza/XIS & 31500 \\
\cite{Sonbas:15} & Draco & \xmm/EPIC & 87\\
\cite{Riemer-Sorensen:15} & 1E~0657-56 (``Bullet cluster'') & NuSTAR & 266 \\
\cite{Jeltema:15} & Draco & \xmm/EPIC & 1660 \\
\bottomrule
\end{tabularx} 
\caption{Summary of searches for dark matter decay line in X-ray observations conducted so far. 
This Table is an update of Table~1 in~\protect\cite{Neronov:13}.
}
\label{tab:bounds-summary} 
\end{table*}

\section*{\sc Observational evidence for the line at $\sim$3.5~keV}

\indent \indent The new emission line at $\sim$3.5~keV is reported by two different 
groups \cite{Bulbul:14a,Boyarsky:14a} in February 2014. 

\begin{figure}[tbh!]
  \centering
  \includegraphics[width=0.99\linewidth]{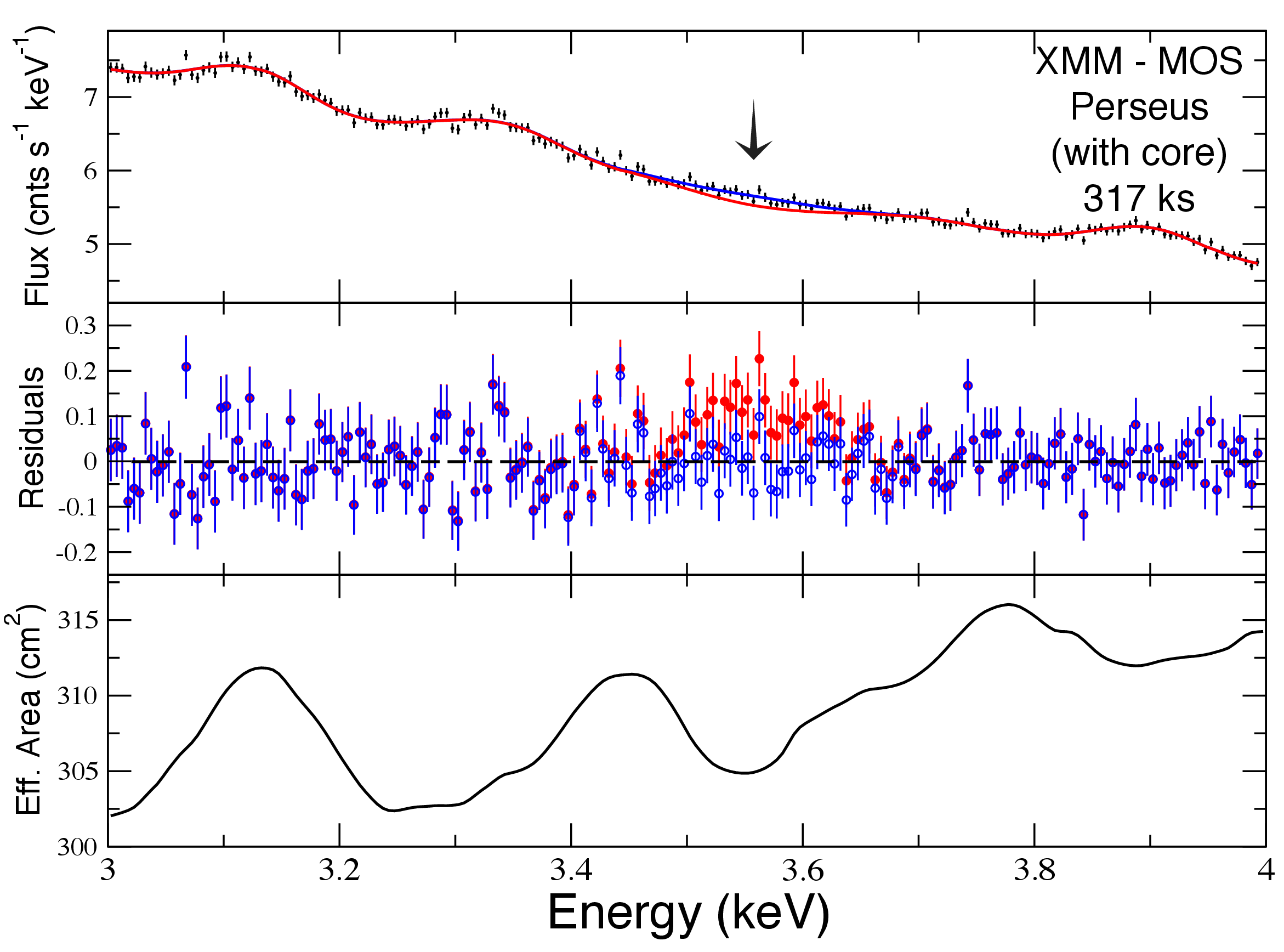}
  \caption{The combined MOS spectrum of Perseus cluster scaled to 3-4~keV energy range. 
  On top of their best-fit model, 
  the series of the single-bin residuals corresponding to the extra emission line at 3.57~keV is shown in red.
  (Adapted from Figure~7 in \cite{Bulbul:14a}).}
  \label{fig:Bulbul-Perseus}
\end{figure}

In \cite{Bulbul:14a}, the authors combine X-ray emission
from the sample of nearby galaxy clusters observed by the European Photon and Imaging Camera 
(EPIC) \cite{Strueder:01,Turner:00} on-board the \xmm\ X-ray cosmic observatory \cite{Jansen:01} with 
the largest number of counts ($> 10^5$ counts for redshifts $z < 0.1$ and $> 10^4$ counts for 
redshifts $0.1 < z < 0.4$). The stacking is made in the cluster's rest frame. As a result, the emission from 
instrumental lines is smeared out, while cosmic lines appear more prominent.  
This method allows \cite{Bulbul:14a} to detect 28 emission lines of astrophysical origin in 2-10~keV band,
much more than in individual galaxy clusters, see e.g. \cite{DePlaa:07}. Apart of them,
\cite{Bulbul:14a} identify the new line located at 3.57$\pm$0.02 
keV in \xmm/MOS \cite{Strueder:01} cameras 
and at 3.51$\pm$0.03 keV in \xmm/PN \cite{Turner:00} camera at the level $\gtrsim 10$ larger than predicted 
by two complexes of nearby astrophysical emission lines located at 3.51~keV (K~XVIII) and 3.62~keV (Ar~XVII). 
The new line is also detected at $> 3\sigma$ local significance in several
different sub-samples of their combined \xmm/EPIC cluster dataset, 
see e.g. Fig.~\ref{fig:Bulbul-Perseus}, and in \chan/ACIS spectrum of Perseus cluster, 
see Table~\ref{tab:line-detections} for details.

\begin{figure}[tbh!]
  \centering
  \includegraphics[width=0.99\linewidth]{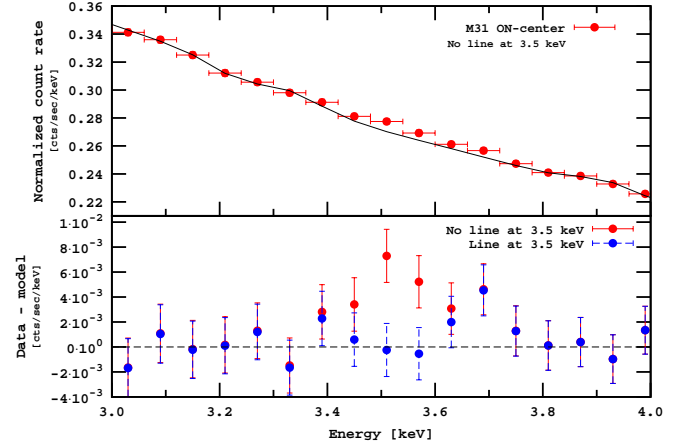}
  \caption{The same as in the previous Figure~\protect\ref{fig:Bulbul-Perseus} but for the combined spectrum of
  Andromeda galaxy.
  (Adapted from Figure~1 in \cite{Boyarsky:14a}).}
  \label{fig:m31_oncen_data_residuals}
\end{figure}

The authors of \cite{Boyarsky:14a} detect the new line at 3.53$\pm$0.03 keV in the central part 
of Andromeda galaxy (see Fig.~\ref{fig:m31_oncen_data_residuals}), 
and in the outskirts of Perseus cluster, see Table~\ref{tab:line-detections}. 
\cite{Boyarsky:14a} exclude the central part of Perseus cluster (analysed in \cite{Bulbul:14a}) 
because of its rather complex structure in X-rays, so the two datasets used in \cite{Boyarsky:14a,Bulbul:14a} 
are totally independent enhancing the statistical significance for the new line. 
Another important result of \cite{Boyarsky:14a}
is the radial dependence of the new line flux in Perseus that appears more consistent with 
decaying dark matter profile than with astrophysical emission. 

\begin{table*}
    \begin{tabularx}{\textwidth}{lXccccc}
      \toprule
 Ref. & Object & Redshift & Instrument & Exposure, & Line position, & Line flux,  \\
      &        &        &              & Ms      & keV            & $10^{-6}$ ph/s/cm$^{2}$ \\
      \otoprule
 \cite{Bulbul:14a} & Full stacked sample & 0.009-0.354 & MOS & 6 & 3.57$\pm$0.02 & 4.0$\pm$0.8 \\
 \cite{Bulbul:14a} & Full stacked sample & 0.009-0.354 & PN & 2 & 3.51$\pm$0.03 & 3.9$^{+0.6}_{-1.0}$ \\
 \cite{Bulbul:14a} & Coma+Centaurus+Ophiuchus & 0.009-0.028 & MOS & 0.5 & 3.57$^a$ & 15.9$^{+3.4}_{-3.8}$ \\
 \cite{Bulbul:14a} & Coma+Centaurus+Ophiuchus & 0.009-0.028 & PN & 0.2 & 3.57$^a$ & $< 9.5$ (90\%) \\
 \cite{Bulbul:14a} & Perseus ($<$ 12') & 0.016 & MOS & 0.3 & 3.57$^a$ & 52.0$^{+24.1}_{-15.2}$ \\
 \cite{Bulbul:14a} & Perseus ($<$ 12') & 0.016 & PN & 0.05 & 3.57$^a$ & $< 17.7$ (90\%) \\
 \cite{Bulbul:14a} & Perseus (1-12') & 0.016 & MOS & 0.3 & 3.57$^a$ & 21.4$^{+7.0}_{-6.3}$ \\
 \cite{Bulbul:14a} & Perseus (1-12') & 0.016 & PN & 0.05 & 3.57$^a$ & $< 16.1$ (90\%) \\ 
 \cite{Bulbul:14a} & Rest of the clusters & 0.012-0.354 & MOS & 4.9 & 3.57$^a$ & 2.1$^{+0.4}_{-0.5}$ \\
 \cite{Bulbul:14a} & Rest of the clusters & 0.012-0.354 & PN & 1.8 & 3.57$^a$ & 2.0$^{+0.3}_{-0.5}$ \\ 
 \cite{Bulbul:14a} & Perseus ($>$ 1') & 0.016 & ACIS-S & 0.9 & 3.56$\pm$0.02 & 10.2$^{+3.7}_{-3.5}$ \\
 \cite{Bulbul:14a} & Perseus ($<$ 9') & 0.016 & ACIS-I & 0.5 & 3.56$^a$ & 18.6$^{+7.8}_{-8.0}$ \\ 
 \cite{Bulbul:14a} & Virgo ($<$ 500'') & 0.003-0.004 & ACIS-I & 0.5 & 3.56$^a$ & $< 9.1$ (90\%) \\
\midrule
 \cite{Boyarsky:14a} & M31 ($<$ 14') & -0.001$^b$ & MOS & 0.5 & 3.53$\pm$0.03 & 4.9$^{+1.6}_{-1.3}$ \\
 \cite{Boyarsky:14a} & M31 (10-80') & -0.001$^b$ & MOS & 0.7 & 3.50-3.56 & $< 1.8$ ($2\sigma$) \\
 \cite{Boyarsky:14a} & Perseus (23-102') & 0.0179$^b$ & MOS & 0.3 & 3.50$\pm$0.04 & 7.0$\pm$2.6 \\
 \cite{Boyarsky:14a} & Perseus (23-102') & 0.0179$^b$ & PN & 0.2 & 3.46$\pm$0.04 & 9.2$\pm$3.1 \\
 \cite{Boyarsky:14a} & Perseus, 1st bin (23-37') & 0.0179$^b$ & MOS & 0.2 & 3.50$^a$ & 13.8$\pm$3.3 \\
 \cite{Boyarsky:14a} & Perseus, 2nd bin (42-54') & 0.0179$^b$ & MOS & 0.1 & 3.50$^a$ & 8.3$\pm$3.4 \\
 \cite{Boyarsky:14a} & Perseus, 3rd bin (68-102') & 0.0179$^b$ & MOS & 0.03 & 3.50$^a$ & 4.6$\pm$4.6 \\
 \cite{Boyarsky:14a} & Blank-sky & --- & MOS & 7.8 & 3.45-3.58 & $< 0.7$ ($2\sigma$) \\ 
\bottomrule
 \end{tabularx}
\caption{Properties of the $\sim$3.5~keV line reported by~\protect\cite{Bulbul:14a,Boyarsky:14a}.
For their analysis, the authors of~\protect\cite{Bulbul:14a,Boyarsky:14a} 
use different X-ray datasets observed by MOS \cite{Turner:00} and PN \cite{Strueder:01} cameras 
on-board \xmm\ observatory \cite{Jansen:01} and ACIS instrument \cite{Garmire:03} 
on-board \chan\ observatory \cite{Weisskopf:00}. 
All error bars are at 1$\sigma$
 (68\%) level.
   \newline
   $^a$ The line position is fixed at given value.
   \newline
   $^b$ The redshift is fixed at NASA Extragalactic Database (NED) value.
   }\label{tab:line-detections}
\end{table*}

The encouraging results of \cite{Bulbul:14a,Boyarsky:14a} have stimulated 
several groups to look on other dark matter-dominated objects. 
The following searches report the presence of the line at $\sim$3.5~keV, 
see Table~\ref{tab:line-confirmation}:
\begin{enumerate}
\item The identification of the line at $\sim$3.5~keV from the region of Galactic 
Centre \cite{Riemer-Sorensen:14,Jeltema:14a,Boyarsky:14b,Carlson:14}. Although it is unclear whether the 
detected line has astrophysical origin
(see the next Sec.~\ref{sec:conventional} for detailed discussion), 
its explanation in terms of decaying dark matter is consistent with the previous new line detections, 
see \cite{Boyarsky:14b,Lovell:14} for details.
\item The detection of the new line in \suza/XIS observations of Perseus, 
Coma and Ophiuchus galaxy clusters \cite{Urban:14}. While subsequent study of \suza/XIS spectra
by \cite{Tamura:14} does not reveal the new line at $\sim$3.5~keV in the central part of Perseus cluster, 
another recent study by \cite{Franse:16} does; however, its apparent discrepancy 
with the negative result of \cite{Tamura:14} is still unclear and has to be resolved further.
\item The detection of the new line at 3.52$\pm$0.08~keV observed 
in X-ray spectra of 8 individual nearby galaxy clusters including Perseus and Coma \cite{Iakubovskyi:15b}.
\end{enumerate}

\begin{figure}[tbh!]
  \centering
  \includegraphics[width=0.99\linewidth]{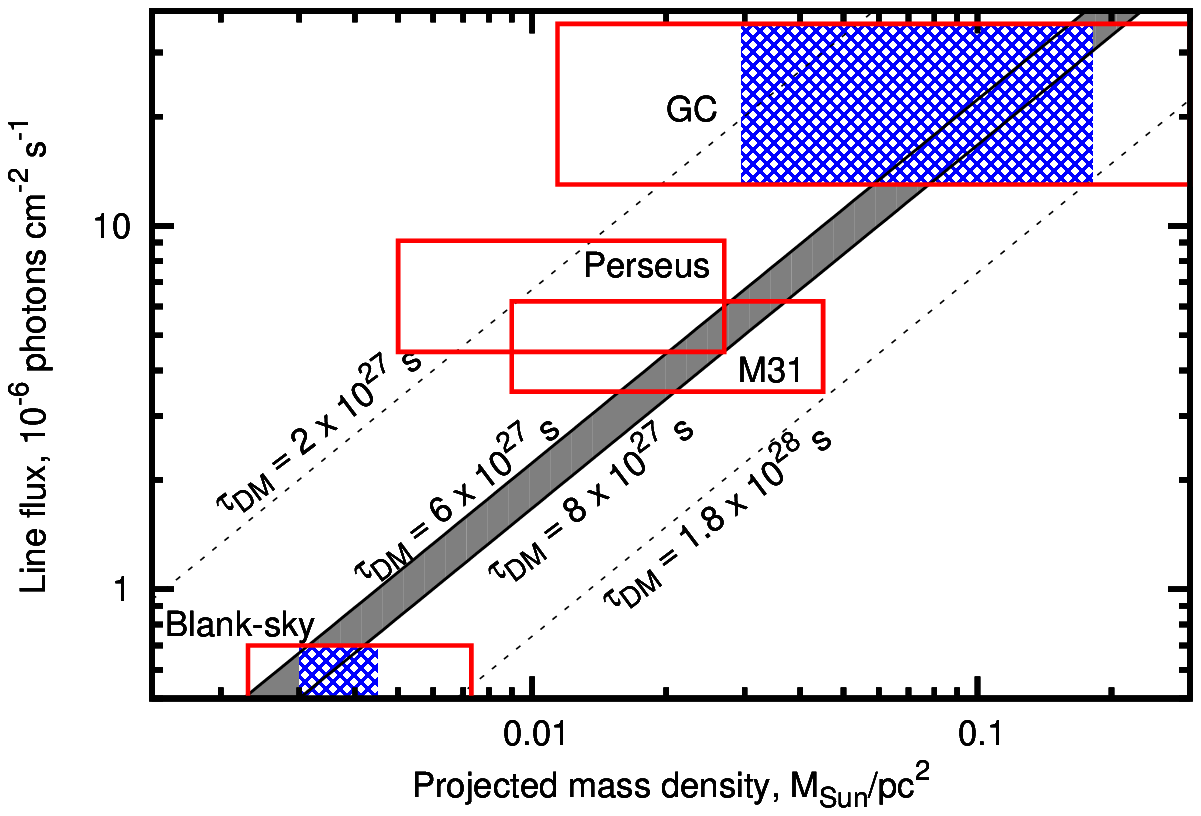}
  \caption{The flux of the $\sim$3.5~keV line from the Galactic Centre, 
  the Perseus cluster outskirts, the Andromeda galaxy, and the `blank sky' dataset ~\protect\cite{Boyarsky:14a} 
  as a function of dark matter projected mass. Diagonal lines show the expected behaviour of
    decaying dark matter signal for a given dark matter particle lifetime.  The vertical sizes
    of the boxes are $\pm 1 \sigma$ statistical error on the line's flux -- or
    the $2\sigma$ upper bound for the blank-sky dataset. The blue shaded regions show a particular 
    Navarro-Frenk-White \cite{Navarro:95,Navarro:96} profile of the Milky Way~\protect\cite{Smith:06}, 
    its horizontal size indicates uncertainties in galactic disk modelling. 
    The lifetime $\tau_\dm \sim (6-8)\times 10^{27}$~s is consistent with all datasets.
    New results from a prolonged Draco \xmm/EPIC observation \cite{Jeltema:15,Ruchayskiy:15}
    give controversial results: while \cite{Jeltema:15} reports an exclusion of dark matter hypothesis
    at $99$\% level, the results of \cite{Ruchayskiy:15} claim that the values of 
    $\tau_\dm \simeq (7-9)\times 10^{27}$~sec
    are still consistent with \emph{all} existing observations.
  (Adapted from Figure~2 in \cite{Boyarsky:14b}).}
  \label{fig:flux-vs-Mproj-blank-final-Smith-wide-line}
\end{figure}

\begin{figure}[tbh!]
  \centering
  \includegraphics[width=0.99\linewidth]{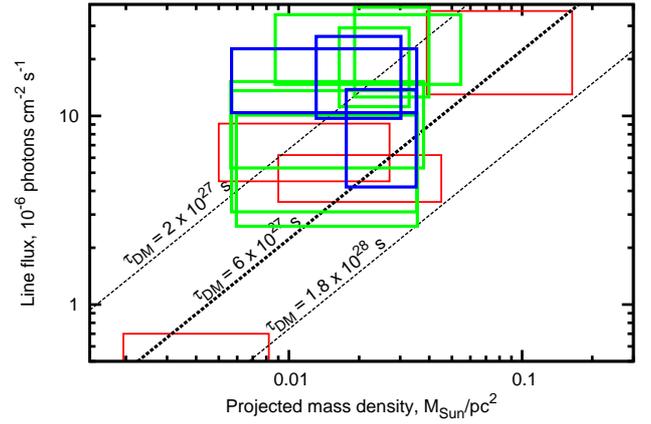}
  \caption{The same as in previous Fig.~\protect\ref{fig:flux-vs-Mproj-blank-final-Smith-wide-line} 
  but over-plotted are the ranges for the $> 2\sigma$ detections in MOS (green) and PN 
  (magenta) cameras, see \cite{Iakubovskyi:15b}. 
  (Adapted from Figure~2 in \cite{Iakubovskyi:15b}).}
  \label{fig:flux-vs-Mproj-blank-v2-Weber-05sigma-newobjects}
\end{figure}

\begin{table*}
    \begin{tabularx}{\textwidth}{lXccccc}
      \toprule
 Ref. & Object & Redshift & Instrument & Exposure, & Line position, & Line flux,  \\
      &        &        &              & Ms      & keV            & $10^{-6}$ ph/s/cm$^{2}$ \\
      \otoprule
 \cite{Riemer-Sorensen:14} & Galactic centre (2.5-12') & 0.0 & ACIS-I & 0.8 & $3.51$ & $\simeq 10^a$ \\ 
\midrule
 \cite{Jeltema:14a} & Galactic centre (0.3-15') & 0.0 & MOS & 0.7 & $3.51$ & $45\pm 4^a$ \\ 
 \cite{Jeltema:14a} & Galactic centre (0.3-15') & 0.0 & PN & 0.5 & $3.51$ & $39\pm 7^a$ \\ 
 \cite{Jeltema:14a} & M31 & 0.0 & MOS & 0.5 & 3.53$\pm$0.07 & 2.1$\pm$1.5$^c$ \\ 
\midrule
\cite{Boyarsky:14b} & Galactic centre ($<$ 14') & 0.0 & MOS & 0.7 & 3.539$\pm 0.011$ & 29$\pm$5 \\ 
\midrule
\cite{Urban:14} & Perseus core ($<$ 6') & 0.0179$^b$ & XIS & 0.74 & 3.510$^{+0.023}_{-0.008}$ & $32.5^{+3.7}_{-4.3}$ \\ 
\cite{Urban:14} & Perseus confined (6-12.7') & 0.0179$^b$ & XIS & 0.74 & 3.510$^{+0.023}_{-0.008}$ & $32.5^{+3.7}_{-4.3}$ \\ 
\cite{Urban:14} & Coma ($<$ 12.7') & 0.0231$^b$ & XIS & 0.164 & $\simeq 3.45^d$ & $\simeq 30^d$ \\ 
\cite{Urban:14} & Ophiuchus ($<$ 12.7') & 0.0280$^b$ & XIS & 0.083 & $\simeq 3.45^d$ & $\simeq 40^d$ \\ 
\cite{Urban:14} & Virgo ($<$ 12.7') & 0.0036$^b$ & XIS & 0.09 & 3.55$^a$ & $< 6.5$ (2$\sigma$) \\ 
\midrule
\cite{Iakubovskyi:15b} & Abell~85 ($<$ 14') & 0.0551$^b$ & MOS & 0.20 & 3.44$^{+0.06}_{-0.05}$ & 6.3$^{+3.9}_{-3.6}$ \\
\cite{Iakubovskyi:15b} & Abell~2199 ($<$ 14') & 0.0302$^b$ & MOS & 0.13 & 3.41$^{+0.04}_{-0.04}$ & 10.1$^{+5.1}_{-4.8}$ \\
\cite{Iakubovskyi:15b} & Abell~496 ($<$ 14') & 0.0329$^b$ & MOS & 0.13 & 3.55$^{+0.06}_{-0.09}$ & 7.5$^{+6.1}_{-4.4}$ \\
\cite{Iakubovskyi:15b} & Abell~496 ($<$ 14') & 0.0329$^b$ & PN & 0.08 & 3.45$^{+0.04}_{-0.03}$ & 16.8$^{+5.9}_{-6.4}$ \\
\cite{Iakubovskyi:15b} & Abell~3266 ($<$ 14') & 0.0589$^b$ & PN & 0.06 & 3.53$^{+0.04}_{-0.06}$ & 8.7$^{+5.1}_{-4.5}$ \\
\cite{Iakubovskyi:15b} & Abell~S805 ($<$ 14') & 0.0139$^b$ & PN & 0.01 & 3.63$^{+0.05}_{-0.06}$ & 17.1$^{+9.3}_{-7.4}$ \\
\cite{Iakubovskyi:15b} & Coma ($<$ 14') & 0.0231$^b$ & MOS & 0.17 & 3.49$^{+0.04}_{-0.05}$ & 23.7$^{+10.7}_{-9.0}$ \\
\cite{Iakubovskyi:15b} & Abell~2319 ($<$ 14') & 0.0557$^b$ & MOS & 0.08 & 3.59$^{+0.05}_{-0.06}$ & 18.6$^{+10.7}_{-7.4}$ \\
\cite{Iakubovskyi:15b} & Perseus ($<$ 14') & 0.0179$^b$ & MOS & 0.16 & 3.58$^{+0.05}_{-0.08}$ & 25.2$^{+12.5}_{-12.6}$ \\
\cite{Iakubovskyi:15b} & Virgo$^e$ ($<$ 14') & 0.0036$^b$ & PN & 0.06 & --- & $< 9.3$ \\
\midrule
\cite{Ruchayskiy:15} & Draco ($<$ 14') & 0.0 & PN & 0.65 & 3.54$^{+0.06}_{-0.05}$ & 1.65$^{+0.67}_{-0.70}$ \\
\midrule
\cite{Franse:16} & Perseus ($<$ 8.3') & 0.0179$^b$ & XIS & 1.67 & 3.54$\pm$0.01 & 27.9$^{+3.5}_{-3.5}$ \\
\cite{Franse:16} & Perseus ($<$ 2') & 0.0179$^b$ & XIS & 1.67 & 3.51$\pm$0.02 & 9.3$^{+2.6}_{-2.7}$ \\
\cite{Franse:16} & Perseus (2'-4.5') & 0.0179$^b$ & XIS & 1.67 & 3.55$\pm$0.02 & 16.7$^{+2.9}_{-3.0}$ \\
\cite{Franse:16} & Perseus (4.5'-8.3') & 0.0179$^b$ & XIS & 1.67 & 3.58$\pm$0.02 & 16.1$^{+3.2}_{-3.4}$ \\
\midrule
\cite{Bulbul:16} & Stacked clusters & 0.01-0.45 & XIS & 8.1 & 3.54$^f$ & 1.0$^{+0.5}_{-0.5}$ \\
\bottomrule
 \end{tabularx}
\caption{Properties of $\sim$3.5~keV line searched after February 2014 in different X-ray datasets 
observed by MOS \cite{Turner:00} and PN \cite{Strueder:01} cameras on-board \xmm\ observatory \cite{Jansen:01}, 
ACIS \cite{Garmire:03} instrument on-board \chan\ observatory \cite{Weisskopf:00} 
and XIS instrument \cite{Koyama:07} on-board \suza\ observatory \cite{Mitsuda:07}.
All error bars are at 1$\sigma$ (68\%) level. 
\newline
$^a$ Best-fit line flux at fixed position 3.51~keV coinciding with the brightest K~XVIII line.
\newline
$^b$ Redshift was fixed at NASA Extragalactic Database (NED) value.
\newline
$^c$ The line is detected at $< 90$\% confidence level. Such a low flux (compared with \cite{Boyarsky:14a}) 
is because of non-physically enhanced level of continuum at 
3-4~keV band used in \cite{Jeltema:14a}, see \cite{Boyarsky:14c} for details.
\newline
$^d$ Parameters estimated from Fig.~3 of \cite{Urban:14}. 
\newline
$^e$ Given an example of the new line non-detection, see Table~II of \cite{Iakubovskyi:15b} for more details.
\newline
$^f$ Line position is fixed at the best-fit energy detected in \suza\ observations of the Perseus 
cluster by~\cite{Franse:16}.
}\label{tab:line-confirmation}
\end{table*}

In summary, positive detections of the new line listed in Table~\ref{tab:line-detections} and 
Table~\ref{tab:line-confirmation} support the hypothesis of decaying dark matter implying radiatively decaying
dark matter lifetime $\tau_\dm \simeq (6-8) \times 10^{27}$~s \cite{Boyarsky:14b,Iakubovskyi:15b,Ruchayskiy:15}.

On the contrary, the following studies do not detect the $\sim$3.5~keV line putting 
the upper bounds on its flux:
\begin{enumerate}
 \item The central part of the Virgo cluster observed by \chan/ACIS \cite{Bulbul:14a},
 \suza/XIS\cite{Urban:14} and \textit{XMM}-\textit{New\-ton}/EPIC \cite{Iakubovskyi:15b},
 as well as other 10 galaxy clusters from \cite{Iakubovskyi:15b}.
 \item Combined spectrum from dwarf spheroidal galaxies \cite{Malyshev:14}.
 \item Outskirts of galaxies \cite{Iakubovskyi:13,Boyarsky:14a,Anderson:14}.
 \item Combined blank-sky observations \cite{Boyarsky:14a,Sekiya:15}.
 \item Prolonged \xmm/EPIC observations of Draco dwarf spheroidal galaxy \cite{Jeltema:15,Ruchayskiy:15};
 although the authors of \cite{Ruchayskiy:15} report a line-like excess at 3.54$\pm$0.06~keV with $\Delta\chi^2 = 5.3$ in PN 
 camera, see Table~\ref{tab:line-detections}, this finding is not supported by an independent analysis 
 of \cite{Jeltema:15} and is not accompanied with similar excess in Draco spectra seen by MOS 
 camera \cite{Ruchayskiy:15,Jeltema:15}.
 \item Combined dataset of 33 galaxy clusters observed by \chan/ACIS~\cite{Hofmann:16}.
\end{enumerate}

At the moment, it is unclear whether these negative searches rule out the decaying dark matter hypothesis 
of the new line. While the bounds obtained in \cite{Malyshev:14} are mildly consistent with the decaying dark 
matter origin of the detections in \cite{Bulbul:14a,Boyarsky:14a}, the results of \cite{Anderson:14} formally 
exclude the decaying dark matter hypothesis of the $\sim$3.5~keV line imposing the very strict $3\sigma$ bound, 
$\tau_\dm > 1.8 \times 10^{28}$~s. Taking into account 
systematic effects in spectra (e.g. causing significant negative residuals) obtained by \cite{Anderson:14} 
and the apparent uncertainty in their dark matter distributions \cite{Boyarsky:09b} would result in 
much weaker bound, see e.g. $\tau_\dm \gtrsim 3.5\times 10^{27}$~s \cite{Iakubovskyi:14} 
using the stacked dataset of nearby galaxies 
of \cite{Iakubovskyi:13} with comparable exposure, still consistent with existing detections.
The uncertainty in dark matter distributions also helps to reconcile the results of the other 
negative searches \cite{Watson:11,Horiuchi:13,Sekiya:15} with $\sim$3.5~keV line detections using
the decaying dark matter paradigm. 
There is also no clarity with the new prolonged ($\sim 1.4$~Ms) \xmm/EPIC observation of Draco 
dwarf spheroidal galaxy -- the object having both well-measured dark matter 
distribution \cite{Geringer-Sameth:14} and proven low X-ray 
background \cite{Jeltema:08,Riemer-Sorensen:09,Malyshev:14,Lovell:14}. 
While \cite{Jeltema:15} reports an exclusion of dark matter hypothesis
at $99$\% level having $2\sigma$ upper bound on radiative dark matter decay lifetime
$\tau_\dm > 2.7\times 10^{28}$~s, the results of \cite{Ruchayskiy:15} 
suggest $\tau_\dm \simeq (7-9)\times 10^{27}$~s, the value still compatible with all existing observations.

\section*{\sc ``Standard'' explanations of the line at $\sim$3.5~keV}\label{sec:conventional}

\indent \indent There are three possible ``standard'' explanations of the new line detections at $\sim$3.5~keV:
\begin{enumerate}
 \item statistical fluctuations;
 \item general-type systematic effects;
 \item astrophysical emission line.
\end{enumerate}

\begin{figure}
  \centering
  \includegraphics[width=0.99\linewidth]{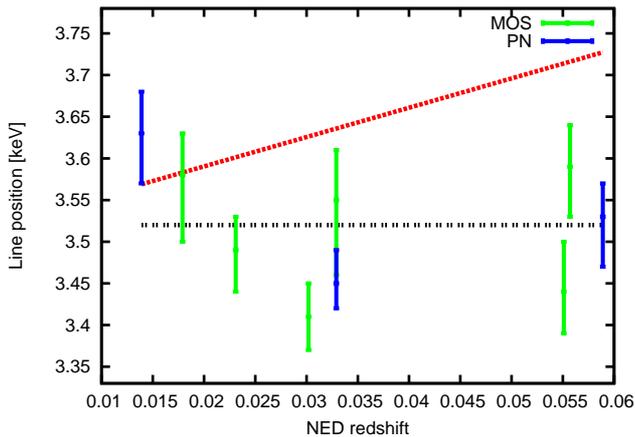}
  \caption{The position of new line detected in \cite{Iakubovskyi:15b} (in the frame of emitting galaxy cluster) 
  as a function of cluster redshift. 
  The red and black dashed lines show the expected behaviour in case of purely systematic 
  and cosmic line origins (assuming the line position 3.52~keV 
  in the detector frame expected from~\protect\cite{Boyarsky:14a,Boyarsky:14b}), respectively.
  (Adapted from Figure~3 in \cite{Iakubovskyi:15b}).}
  \label{fig:line-position-z}
\end{figure}

With recent increase of positive detections reported by \cite{Iakubovskyi:15b}, it is very hard 
to explain \emph{all} of the detections with pure statistical fluctuations. Nevertheless, statistical 
fluctuations may be responsible for new line detections or non-detections in \emph{some} individual objects,
as well as for variations of the detected line position up to $\sim$110~eV \cite{Iakubovskyi:15b}, see 
Fig.~\ref{fig:line-position-z} -- the effect that should be properly taken into account when searching for 
the new line (unlike \cite{Malyshev:14,Anderson:14,Urban:14}).

The systematic origin of the line is carefully investigated because of the previous study 
of the line-like residual at $\sim$2.5~keV in the Willman~1 dwarf spheroidal, 
see Sec.~\ref{sec:previous-searches} for details.
However, the explanation of the $\sim$3.5~keV line with general-type systematics 
suggested in \cite{Jeltema:14a} is unlikely.
For example, its position (in the frame of emitting object) 
remains remarkably constant with redshift \cite{Bulbul:14a,Boyarsky:14a,Iakubovskyi:15b},
see Fig.~\ref{fig:line-position-z}, which cannot be explained by simple systematics.
The line is also independently detected by five detectors on-board three cosmic missions, see
Table~\ref{tab:line-detections} and Table~\ref{tab:line-confirmation}.
Finally, similar feature of systematic origin should be detected in the blank-sky dataset \cite{Boyarsky:14a}, 
and should have different radial behaviour in the outskirts of Perseus cluster \cite{Boyarsky:14a,Franse:16}.

On the other hand, the explanation of the new line with the K~XVIII line complex at $\sim$3.5~keV 
suggested by \cite{Jeltema:14a}
(see also an extensive discussion in \cite{Bulbul:14a,Boyarsky:14c,Bulbul:14b,Jeltema:14b}) is still 
possible, at least for Galactic Centre region and galaxy clusters, contrary to initial claims 
of \cite{Bulbul:14a,Boyarsky:14a}. The reason is that the emission flux from the K~XVIII line complex at 
$\sim$3.51~keV suggested by \cite{Jeltema:14a} is highly uncertain due to large 
uncertainties of the Potassium abundance, see e.g. \cite{Romano:10,Phillips:15} for a 
potential\footnote{The results of \cite{Phillips:15} indicate an order of magnitude over-abundance
of Potassium in solar corona compared to solar photosphere. Based on this result, \cite{Phillips:15}
suggested that the Potassium abundance in hot plasma in galaxies and galaxy clusters may have also been 
enhanced compared to the solar photospheric values. However, because at the moment there is no 
established mechanism that could effectively provide such an enhancement, the results of \cite{Phillips:15}
only indicate the \emph{potential} level of uncertainty, similar to the measurements in \cite{Romano:10}.} 
level of uncertainty. 
Moreover, unlike other possible emission lines of astrophysical origin near $\sim$3.5~keV 
(such as Cl~XVII lines at 3.51~keV found largely 
sub-dominant in Galactic Centre region \cite{Jeltema:14a} and in galaxy clusters \cite{Bulbul:14b}), 
K~XVIII line complex
does not have stronger counterparts at other energies and can hardly be excluded by measurements of other lines,
the strongest of them is the K~XIX line complex at 3.71~keV of comparable strength \cite{Iakubovskyi:15a}.
The same is true about the charge exchange of S~XVI ions recently suggested by \cite{Gu:15}.

An alternative approach is to study the \emph{line morphology}. 
At the moment, two different methods have been used. The first method \cite{Boyarsky:14a,Bulbul:14a} 
is to split the region covered by 
astrophysical sources onto several independent subregions, 
large enough to detect the line at in each of them, and to model their spectra separately looking for a 
line-like excess in each of them.
As a result, \cite{Bulbul:14a} show that the $\sim$3.5~keV line in Perseus cluster is somewhat more 
concentrated compared to decayed dark matter distributed according to 
Navarro-Frenk-White \cite{Navarro:95,Navarro:96} profile. 
By studying the $\sim$3.5~keV line emission from Perseus cluster outskirts, \cite{Boyarsky:14a} obtain
that such distribution is better consistent with radiatively decaying dark matter distributed according to 
the well-established Navarro-Frenk-White profile than with astrophysical continuum emission distributed 
according to the isothermal $\beta$-model of \cite{Cavaliere:76}.
The recent detailed study \cite{Franse:16} confirms this result and expands it to the central 
region of Perseus cluster. 

The second method to study the line morphology \cite{Carlson:14} 
deals with spatial distribution of the `line plus continuum' X-ray emission 
in Perseus cluster and Galactic Centre region 
with further eliminating continuum component by either assuming it spatially smooth or 
cross-correlating the `line plus continuum' images in several energy bands 
(including those dominated by astrophysical line emission).
By using the second method, the authors of \cite{Carlson:14} show that adding decaying dark matter distribution 
from a \emph{smooth} dark matter profile (Navarro-Frenk-White, Einasto, Burkert) 
does not improve the fit quality in both objects, and demonstrate that distribution of the events in 
3.45-3.6~keV bands correlates with that in the energy bands of strong astrophysical emission, rather than 
with that in line-free energy bands. 
Based on these findings, Ref. \cite{Carlson:14} claims the exclusion of decaying 
dark matter origin of 3.5~keV in Galactic Centre and Perseus cluster.

\begin{figure}
  \centering
  \includegraphics[width=0.99\linewidth]{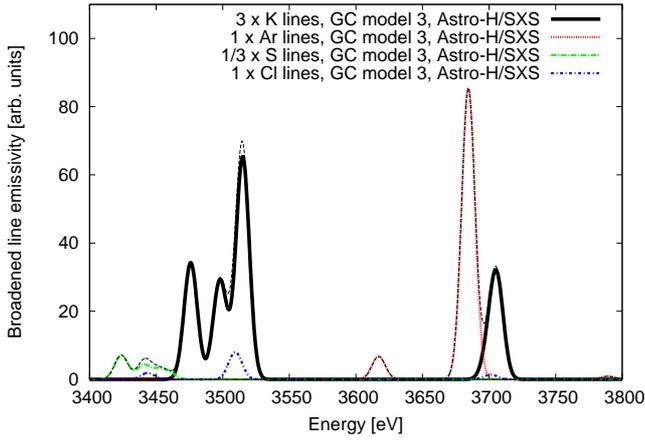}
  \caption{Line emissivities (in arbitrary units) broadened with energy resolution of Soft X-ray Spectrometer 
  (SXS) on-board  \textit{Hitomi} (former \textit{Astro-H}),  
  $\sigma_{SXS} = 5$~eV, as functions of energy for three-component model of~\protect\cite{Jeltema:14a} of 
  Galactic Centre. The relative S, Ar, Cl and K abundances are set to 1/3 : 1 : 1 : 3, according to 
  Sec.~2.2 of~\protect\cite{Jeltema:14a}. Thin dashed line shows the total line emissivity. 
  (Adapted from Figure~2 in \cite{Iakubovskyi:15a}).}
  \label{fig:K-lines-3400-3800-GC3-sxs}
\end{figure}

To ultimately check the astrophysical origin of the $\sim$3.5~keV line, new observations with high-resolution
imaging \footnote{\emph{Grating} spectrometers such as \chan/HETGS \cite{Canizares:05} have excellent spectral 
resolution for \emph{point} sources; however, for extended ($\gtrsim$1~arcmin) sources their spectral 
resolution usually degrades to that for existing imaging spectrometers, see e.g. \cite{Dewey:02}.} 
spectrometers such as Soft X-ray Spectrometer (SXS) \cite{Mitsuda:14}
on-board the recently launched \emph{Hitomi}\footnote{Although
\emph{Hitomi} is now broken apart, 
it had observed Perseus cluster before the break-up \cite{Kelley:16,Hitomi-Perseus-Nature}.}
(former \emph{Astro-H}) 
mission \cite{Takahashi:14}, \textit{Micro-X} 
sounding rocket experiment \cite{Figueroa-Feliciano:15} and the X-ray Integral Field Unit 
(X-IFU) \cite{Barret:13,Ravera:14} on-board planned \textit{Athena} mission \cite{Nandra:13,Barret:13conf}.
If the position of the new line incidentally coincides with that of K~XVIII line complex, more detailed study 
of the ratios of the Potassium line emissivities will be essential to finally check the 
astrophysical origin of the new line, see Fig.~\ref{fig:K-lines-3400-3800-GC3-sxs} for details. 

\section*{\sc Other extra line candidates in X-ray range}

\indent \indent Although the line at $\sim$3.5~keV receives the largest attention of the community,
there are three other line candidates in X-rays which origin is also not established:

\begin{enumerate}
\item According to \cite{Prokhorov:10}, intensity of the \emph{Fe XXVI Ly-$\gamma$ line} at 8.7~keV
observed in \suza/XIS spectrum of the Milky Way centre \cite{Koyama:06} cannot be
explained by standard ionization and recombination processes, and
dark matter decay may be a possible explanation of this excess.
\item According to Sec.~1.4 of \cite{Koyama:14}, two faint extra line-like excesses at 9.4 and 10.1~keV are 
detected in the combined \suza/XIS spectrum of Galactic Bulge region. Notably, positions of these excesses do not 
coincide with any bright\footnote{The newest available atomic database AtomDB v.3.0.2 \cite{Foster:14} 
contains several faint Ni~XXVI - Ni~XXVIII emission lines at 10.02-10.11~keV.} 
astrophysical or instrumental line and their intensities can be explained in frames of decaying dark matter
origin, see right Fig.~8 of \cite{Koyama:14}.
\end{enumerate}

\section*{\sc Possible implications for new physics}\label{sec:implications}

\indent \indent If none of ``conventional'' explanations discussed in the previous Sec.~\ref{sec:conventional} were valid,  
the existence of the new line at $\sim$3.55~keV will be an indication of a new physics beyond the Standard Model.

\begin{figure}[!t]
  \centering
  \includegraphics[width=\linewidth]{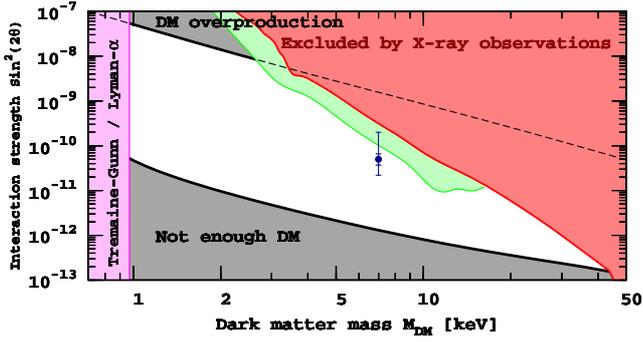}
  \caption{Constraints on sterile neutrino dark matter within the $\nu$MSM 
  model \cite{Asaka:05a,Asaka:05b,Boyarsky:09a}. 
In every point in the white region sterile neutrinos constitute 100\% of dark matter and their properties 
agree with the existing bounds. 
The blue point corresponds to the observed line from Andromeda galaxy,
while the error bars indicate statistical errors (thick) and uncertainty in dark matter distribution
at the central part of Andromeda galaxy (thin). 
(Adapted from Figure~4 in \cite{Boyarsky:14a}).}
\label{fig:numsm}
\end{figure}

Historically, the first model discussed in connection with $\sim$3.5~keV detection is the neutrino minimal
extension of the Standard Model with three right-handed (sterile) neutrinos 
(the \numsm) \cite{Asaka:05a,Asaka:05b,Boyarsky:09a}. In this model, the lightest sterile neutrino with 
mass in keV range forms the bulk of dark matter while two heavier sterile neutrinos are responsible for two other
established phenomena beyond the Standard Model -- neutrino oscillations and generation of asymmetry between 
baryons and anti-baryons in early Universe. Sterile neutrinos decay possess the 2-body radiative channel
$N \to \gamma + \nu$, so the observation of $\sim$3.5~keV decay line would imply the existence of light sterile 
neutrino dark matter particles with mass $\sim$7.1~keV. The simplest production scenario of sterile neutrino 
dark matter -- via non-resonant oscillations of usual (active) neutrinos in the early 
Universe \cite{Dodelson:93,Dolgov:00,Abazajian:01a,Abazajian:01b,Abazajian:05,Asaka:06b} --
is already excluded by the combination of X-ray measurements \cite{Boyarsky:07a}, measurements of Lyman-$\alpha$ 
forest \cite{Viel:05,Viel:06,Seljak:06,Viel:07,Boyarsky:08c,Boyarsky:08d} and the 
phase-space bound from dwarf spheroidal 
galaxies \cite{Tremaine:79,Boyarsky:08a,Gorbunov:08b,Angus:09,Shao:12}. The realistic 
scenario of dark matter production within the \numsm\ now involves resonant oscillations of active neutrinos 
in hot primeval plasma with significant lepton asymmetry generated by decays of heavier sterile 
neutrinos \cite{Shi:98,Laine:08a,Abazajian:14,Ghiglieri:15,Venumadhav:15}.
The parameters of observed $\sim$3.5~keV line are consistent with \numsm\ predictions, 
see Fig.~\ref{fig:numsm} for details.
Because the interaction of sterile neutrino dark matter with Standard Model particles is orders of magnitude 
weaker than that of ordinary neutrinos, its prospects for direct detection in a particle physics experiment
are very far from the existing experimental technique, 
see \cite{Liao:10,Liao:13,Mertens:14,Dragoun:15,Adhikari:16}. 
To confirm the \numsm, a search for heavier sterile neutrinos in GeV range is needed, handled by e.g. 
planned Search for Hidden Particles (\textit{SHiP}) experiment \cite{Bonivento:13,Alekhin:15} and 
Future electron-positron e$^+$e$^-$ Circular Collider (\textit{FCC-ee}) \cite{Blondel:14}.

However, the confirmation of decaying dark matter origin of the new line 
does not imply the existence of \numsm\ 
sterile neutrinos as there are plenty of other alternatives which can potentially explain the 
$\sim$3.55~keV line, see e.g. \cite{Boyarsky:14b,Iakubovskyi:14,Adhikari:16} and the references therein.
Differences among these models can be further probed by:
\begin{itemize}
 \item changes in the new line morphology because of non-negli\-gible initial dark matter velocities, see 
 e.g. \cite{Maccio:12b,Lovell:13a};
 \item other astrophysical and cosmological tests, 
 see e.g. \cite{Boyarsky:08a,Viel:13,Abazajian:14,Schneider:14,Merle:14,Bozek:15,Lovell:15,Bose:15,
 Horiuchi:15,Wang:15,Li:15,Bose:16a,Ludlow:16,Schneider:16,Kamada:16,Rudakovskyi:16,Bose:16b};
 \item search for ``smoking gun'' signatures in future dedicated particle physics 
 experiments, such as \textit{SHiP} \cite{Bonivento:13,Alekhin:15} and 
 \textit{FCC-ee} \cite{Blondel:14} experiments.
\end{itemize}

Recently proposed alternatives to radiatively decaying dark matter include: decay of excited dark 
matter states \cite{Finkbeiner:14,Cline:14,Okada:14b,Cline:14b,Boddy:14,Schutz:14,Cline:14c,Berlin:15,DEramo:16}, 
annihilating dark matter \cite{Dudas:14,Frandsen:14,Baek:14b,Mambrini:15}, 
dark matter decaying into axion-like particles with further conversion to photons in magnetic 
field \cite{Cicoli:14,Conlon:14a,Conlon:14b,Alvarez:14,Berg:16}. 
These models predict \emph{substantial} difference in 
$\sim$3.5~keV line morphology compared to the radiatively decaying dark matter. For example, the spatial 
distributions of the new line in these models should be more concentrated towards the centres of 
dark matter-dominated objects compared to radiatively decaying dark matter, e.g. due to
larger dark matter density (for excited and annihilating dark matter) or larger magnetic fields 
(for magnetic field conversion of axion-like particles). Further non-observation of the $\sim$3.5~keV line 
in outskirts of dark matter-dominated objects would argue in favour of these models.

\section*{\sc Conclusion and future directions}

\indent \indent The origin of the new emission line at $\sim$3.5~keV reported 
by \cite{Bulbul:14a,Boyarsky:14a,Boyarsky:14b,Urban:14,Iakubovskyi:15b} remains unexplained. 
The observed properties of the new line are consistent with radiatively decaying dark matter and other 
interesting scenarios (such as, exciting dark matter, annihilating dark matter and dark matter decaying 
into axion-like particles further converted in cosmic magnetic fields) motivated by various particle physics 
extensions of the Standard Model. In case of radiatively decaying dark matter, further detections 
would lead to direct detection of new physics. Specially 
dedicated observations using existing X-ray missions (such as \xmm, \chan, \suza) 
still allow such detections although one should take detailed care on various 
systematic effects that could mimic or hide the new line. 

The alternative is to use new better instruments. The basic requirements for such instruments -- higher 
\emph{grasp} (the product of field-of-view and effective area) and better \emph{spectral resolution} -- 
have first formulated in \cite{Boyarsky:06f}. Both the soft X-ray Spectrometer \cite{Mitsuda:14} on-board 
the new X-ray mission \textit{Hitomi} (former \textit{Astro-H}) \cite{Takahashi:14,Kitayama:14}
and the plan\-ned \textit{Micro-X} 
sounding rocket experiment \cite{Figueroa-Feliciano:15} meet only second requirement 
having the energy resolution by an order of magnitude better ($\sim 5$~eV) than existing imaging spectrometers.
Before being broken apart, \textit{Hitomi} has already 
observed Perseus cluster \cite{Kelley:16}. 
It is expected \cite{Bulbul:14a} that such an observation would allow \textit{Hitomi}
to precisely determine the new line position in brightest objects with prolonged observations
and to detect the K~XIX emission line complex at $\sim$3.71~keV. 
Another possible option is to resolve the intrinsic width 
of the new line because of its Doppler broadening in 
galaxies and galaxy clusters \cite{Bulbul:14a,Speckhard:15}.
As a result, 
\textit{Hitomi}/SXS is able to check whether 
the new line comes from new physics or from 
(anomalously enhanced) astrophysical emission. 
The same is expected from the \textit{Micro-X} rocket-based micro-calorimeter 
(to be launched in 2017) which will 
observe the central region of our Galaxy. 
Another possibility is to use the planned \textit{eROSITA} instrument on-board 
\textit{Spektrum-R\"{o}ntgen-Gamma} mission \cite{Merloni:12} and 
the planned \textit{LOFT} mission \cite{Zane:14} 
which high grasp and moderate energy resolution would allow to detect the new line at much smaller 
intensities \cite{Neronov:13,Zandanel:15}. 
Finally, an ``ultimate'' imaging spectrometer proposed in e.g. \cite{Boyarsky:12c}
(an example is the X-ray Integral Field Unit (X-IFU) \cite{Barret:13,Ravera:14} on-board the planned 
\textit{Athena} mission \cite{Nandra:13,Barret:13conf})
would reveal the detailed morphology structure of the $\sim$3.5~keV line \cite{Neronov:15}.

\section*{\sc acknowledgement}

\indent \indent This work was supported by a research grant from VILLUM FONDEN. The author also acknowledges
partial support from the Swiss National Science Foundation grant SCOPE IZ7370-152581, 
the Program of Cosmic Research of the National Academy of 
Sciences of Ukraine, the State Fund for Fundamental Research of Ukraine 
and the State Programme of Implementation of Grid Technology in Ukraine during early stages of this work.


\bibliographystyle{naturemag} 
\bibliography{preamble,narev-iakubovskyi}

\end{document}